\documentclass[12pt]{iopart}

\usepackage{graphicx}
\usepackage{hyperref}
\usepackage{epsfig}
\usepackage{verbatim}
\usepackage{cite}

\expandafter\let\csname equation*\endcsname\relax
\expandafter\let\csname endequation*\endcsname\relax

\usepackage{amsmath}
\usepackage{amssymb,amstext}

\newtheorem{theorem}{Theorem}

\begin{document}

\title{Robust Adaptive Quantum Phase Estimation}

\author{Shibdas Roy$^{1,2,*}$, Ian R. Petersen$^{1}$ and Elanor H. Huntington$^{2,3}$}
\address{$^1$School of Engineering and Information Technology, University of New South Wales, Australian Defence Force Academy, Canberra}
\address{$^2$Australian Research Council Centre of Excellence for Quantum Computation and Communication Technology, Australia}
\address{$^3$College of Engineering and Computer Science, Australian National University, Canberra}

\eads{\mailto{$^{*}$roy\_shibdas@yahoo.co.in}}

\date{\today}

\begin{abstract}
Quantum parameter estimation is central to many fields such as quantum computation, communications and metrology. Optimal estimation theory has been instrumental in achieving the best accuracy in quantum parameter estimation, which is possible when we have very precise knowledge of and control over the model. However, uncertainties in key parameters underlying the system are unavoidable and may impact the quality of the estimate. We show here how quantum optical phase estimation of a squeezed state of light exhibits improvement when using a robust fixed-interval smoother designed with uncertainties explicitly introduced in parameters underlying the phase noise.\\

\end{abstract}

\maketitle

\section{Introduction}
Quantum parameter estimation \cite{WM} is the problem of estimating a classical variable of a quantum system. It plays a key role in quantum computation \cite{HWA}, quantum communications \cite{SPK,CHD}, quantum key distribution \cite{IWY}, metrology \cite{GLM1} and gravitational wave interferometry \cite{GMM}, etc. A common and technologically relevant example is estimating an optical phase in a quantum system. Optimal estimation theory has earlier been considered in devising and improving quantum parameter estimation techniques. Systematic approaches to optimal estimation yield estimates with the lowest mean-square estimation error. This has helped achieve better estimation accuracies than otherwise obtained previously \cite{TW,YNW}.

Nonetheless, the optimality of the estimation process relies on precise knowledge of the system model. However, this is usually unrealistic due to inevitable modelling errors. In many cases, it is impossible to precisely measure and determine values of relevant model parameters in an experiment. This is detrimental in problems of quantum estimation because any uncertainty in our knowledge of the parameters in the system model may result in considerable degradation in the estimation accuracy. It is, therefore, desired to make the estimation process robust to uncertainties in the underlying model parameters \cite{LXP,ZDG}.

It is important in many practical engineering problems to ensure that the critical measures of the system performance do not deviate beyond certain thresholds. Such thresholds mark the point beyond which the system has a high risk of breaking down or becoming unusable. In quantum estimation problems, performance is typically determined by the error in the estimation process in the presence of uncertainty in the system. An optimal estimator is optimized by minimizing a cost function to yield the least mean-square estimation error for exact model parameters. When the model parameters are not the same as the true system parameters, i.e. when there is uncertainty in one or more of the model parameters, the true estimation error may be worse than the predicted optimal value. In a worst-case situation with large uncertainties, this increase in the size of the estimation errors could be significant. A robust estimator, on the other hand, can be designed by optimizing the worst case of a cost function for an uncertain system model. This allows the robust estimator to yield lower estimation errors than the optimal estimator in the worst-case scenario. For example, in gravitational wave detection, large estimation errors may mask a gravitational-wave event or imitate an event. Such a false event may be avoided using a robust estimator, which has a sufficient guaranteed worst-case precision.

In this paper, we aim to design a robust estimator for quantum phase estimation that provides guaranteed worst-case performance. Robust quantum parameter estimation was previously considered in Ref. \cite{JS} for magnetometry. That paper employed heuristic feedback mechanism to achieve robustness. By contrast, we consider a more systematic approach to robust estimation in a state-space setting with explicitly modelled uncertainty. Among other related works, Ref. \cite{NY} proposed a robust quantum observer for uncertain linear quantum systems and Ref. \cite{JNP} considered robustness in the context of coherent feedback. However, for linear quantum systems, much of the rich classical estimation theory may be applied. To our knowledge, the potential application of classical robust estimation theory has not yet been explored in improving quantum estimation techniques.

Quantum phase estimation has been area of active research recently \cite{HMW,WK1,WK2,PWL,MA,BW1,BW2,TSL}. Adaptive quantum phase estimation of the continuously varying phase of a coherent state of light using smoothing was demonstrated in Ref. \cite{TW}. Fixed-interval smoothing uses both past and future measurements in a fixed time-interval to yield a more accurate estimate than obtained using only past measurements \cite{MT1,LK,JSM,FP,WWS,DQM,DCF,RKM}. Using a robust fixed-interval smoother \cite{MSP}, the estimation process for the adaptive experiment can be improved in the presence of uncertainty in the underlying phase noise subject to an Ornstein-Uhlenbeck (OU) noise \cite{RPH2}. While a coherent state has the same uncertainty in both (amplitude and phase) quadratures, a squeezed state has reduced fluctuations in one of the two quadratures at the expense of increased fluctuations in the other. Using a squeezed state of light provides quantum enhancement in adaptive phase tracking \cite{YNW}. Here, we illustrate the guaranteed worst-case performance of the robust estimator for such a squeezed state \cite{RPH3}. We model the phase to be estimated as an OU process to begin with in our paper because Refs. \cite{TW} and \cite{YNW} consider such a noise mechanism. The idea is to demonstrate the improvement provided by our robust estimator over the optimal estimator used in the noise setting of Ref. \cite{YNW}. Such a stochastically varying phase resembles a continuous-time random walk with a tendency to return to the mean phase of zero, a kind of noisy relaxation process that occurs in many physical situations, and is more relevant for applications such as physical metrology and communication than a time-invariant (but initially unknown) phase \cite{YNW}.

The robust fixed-interval smoothing scheme can as well be applied to estimate a phase, modelled as a resonant noise process with uncertainty in its parameters \cite{RPH5}. A related robust filtering problem for coherent state was considered by the authors for OU process in Ref. \cite{RPH1} and for resonant process in Ref. \cite{RPH4}. Here, we build on the results in the conference paper \cite{RPH5} to provide interesting insights about the guaranteed worst-case performance of the robust estimator. Moreover, we show that the performance improvement of the robust estimator relative to the optimal estimator grows as the noise process becomes more resonant. We also show here that the worst-case performance of our robust estimator relative to the optimal estimator is better for realistic lossy squeezed beams than that for ideal pure squeezed beams at the optimal degrees of squeezing. In addition, we illustrate that our robust estimator exhibits an optimal photon number for which its relative performance is the best with respect to the optimal estimator.

\section{Optimal Estimator}
The optimal estimator in Ref. \cite{YNW} involves an offline optimal smoother, in addition to a Kalman filter in the feedback loop. The feedback Kalman filter is a causal filter. However, the smoother is acausal, since it is, in principle, a combination of a forward-time Kalman filter and a backward-time Kalman filter, the estimates of which are combined to yield the optimal smoothed estimate \cite{RGB}. While the forward Kalman filter is essentially the feedback filter itself and uses only past measurements, the backward filter yields its estimate based on future measurements with respect to the time of the desired smoothed estimate within the chosen fixed time-interval $[0, \tau]$. A smoother, therefore, cannot be used to produce real-time estimates, and is usually used for offline data processing or with a delay with respect to the estimation time to yield more accurate estimates than obtained using the feedback Kalman filter alone \cite{TW,YNW}.

\subsection{System Model}
We need to define our system in terms of the process and measurement models in a state-space setting.

The process model is the OU noise process that modulates the phase $\phi(t)$, to be estimated, of the continuous optical phase-squeezed beam \cite{YNW}:
\begin{equation}\label{eq:ou_noise}
\dot{\phi}(t) = -\lambda\phi(t) + \sqrt{\kappa}v(t),
\end{equation}
where $\lambda^{-1} > 0$ is the correlation time of $\phi(t)$, $\kappa > 0$ is the phase variation magnitude and $v(t)$ is a zero-mean white Gaussian noise with unity amplitude.

The phase-modulated beam is measured by homodyne detection using a local oscillator, the phase of which is adapted with the filtered estimate $\hat{\phi}_f(t)$ using feedback, thereby yielding a normalized homodyne output photocurrent \cite{YNW}:
\begin{eqnarray}
I(t)dt &\simeq 2|\alpha|[\phi(t)-\hat{\phi}_f(t)]dt + \sqrt{\overline{R}_{sq}}dW(t),\\
\overline{R}_{sq} &= \sigma_f^2 e^{2r_p}+(1-\sigma_f^2)e^{-2r_m},
\label{eq:implicit}
\end{eqnarray}
where $|\alpha|$ is the amplitude of the input phase-squeezed beam, and $W(t)$ is a Wiener process arising from squeezed vacuum fluctuations. The parameter $\overline{R}_{sq}$ is determined by the degree of squeezing ($r_m \geq 0$) and anti-squeezing ($r_p \geq r_m$) and by $\sigma_f^2$ (see later). We use the measurement appropriately scaled as our measurement model \cite{RPH3}:
\begin{equation}\label{eq:ou_measurement}
\theta(t) := \frac{1}{\sqrt{\overline{R}_{sq}}}[I(t)+2|\alpha|\hat{\phi}_f(t)]=\frac{2|\alpha|}{\sqrt{\overline{R}_{sq}}}\phi(t) + w(t),
\end{equation}
where $w := \frac{dW}{dt}$ is also a zero-mean white Gaussian noise with unity amplitude.

Here, $E[v(t)v^T(r)] = {N}\delta(t - r)$, $E[w(t)w^T(r)] = {S}\delta(t - r)$, $E[v(t)w^T(r)] = 0$, where $E[\cdot]$ denotes the expectation value and $\delta(\cdot)$ is the delta function. Since $v$ and $w$ are of unity amplitude, both ${N}$ and ${S}$ are unity.

\subsection{Forward Filter}
For the process and measurement models given by (\ref{eq:ou_noise}) and (\ref{eq:ou_measurement}) respectively, the standard steady-state Kalman filter is constructed by solving a continuous-time algebraic Riccati equation as in \ref{sec:met_opt_smth}.

The steady-state Riccati equation to be solved for the forward Kalman filter is:
\begin{equation}
-2\lambda P_f - \frac{4|\alpha|^2}{\overline{R}_{sq}} P_f^2 + \kappa = 0,
\end{equation}
where $P_f = \sigma_f^2$ is the forward filter error-covariance. The stabilising solution of the above equation is:
\begin{equation}\label{eq:fwd_err_cov_kalman}
{P_f = \frac{\overline{R}_{sq}}{4|\alpha|^2} \left( -\lambda + \sqrt{\lambda^2 + \frac{4\kappa|\alpha|^2}{\overline{R}_{sq}}} \right).}
\end{equation}

The forward filter equation is:
\begin{equation}\label{eq:kalman_fwd_filter}
\dot{\hat{\phi}}_f = -(\lambda +\frac{2|\alpha|K_f}{\sqrt{\overline{R}_{sq}}})\hat{\phi}_f + \frac{2|\alpha|K_f}{\sqrt{\overline{R}_{sq}}}\phi + K_fw,
\end{equation}
where $K_f = \frac{\sqrt{\overline{R}_{sq}}}{2|\alpha|}\left(-\lambda + \sqrt{\lambda^2+\frac{4\kappa|\alpha|^2}{\overline{R}_{sq}}}\right)$ is the forward Kalman gain.

\subsection{Backward Filter}
The steady-state backward Kalman filter is constructed similarly as in \ref{sec:met_opt_smth}.

The steady-state Riccati equation to be solved for the backward Kalman filter is:
\begin{equation}
2\lambda P_b - \frac{4|\alpha|^2}{\overline{R}_{sq}} P_b^2 + \kappa = 0,
\end{equation}
where $P_b = \sigma_b^2$ is the backward filter error-covariance. The stabilising solution of the above equation is:
\begin{equation}\label{eq:bwd_err_cov_kalman}
{P_b = \frac{\overline{R}_{sq}}{4|\alpha|^2} \left( \lambda + \sqrt{\lambda^2 + \frac{4\kappa|\alpha|^2}{\overline{R}_{sq}}} \right).}
\end{equation}

The backward filter equation is:
\begin{equation}\label{eq:kalman_bwd_filter}
\dot{\hat{\phi}}_b = (\lambda - \frac{2|\alpha|K_b}{\sqrt{\overline{R}_{sq}}})\hat{\phi}_b + \frac{2|\alpha|K_b}{\sqrt{\overline{R}_{sq}}}\phi + K_bw,
\end{equation}
where $K_b = \frac{\sqrt{\overline{R}_{sq}}}{2|\alpha|}\left(\lambda + \sqrt{\lambda^2 + \frac{4\kappa|\alpha|^2}{\overline{R}_{sq}}}\right)$ is the backward Kalman gain.

\subsection{Smoother Error}
The smoother error, $P_s = \sigma^2$, is obtained by combining the forward and backward Kalman filter errors as in \ref{sec:met_opt_smth}, i.e.
\begin{equation}\label{eq:opt_smoother_error}
P_s = (P_f^{-1}+P_b^{-1})^{-1},
\end{equation}
since the forward and backward estimates are independent. From (\ref{eq:fwd_err_cov_kalman}), (\ref{eq:bwd_err_cov_kalman}), (\ref{eq:opt_smoother_error}), we get:
\begin{equation}
P_s = \frac{\kappa}{2\sqrt{\lambda^2+\frac{4\kappa|\alpha|^2}{\overline{R}_{sq}}}},
\end{equation}
which matches with Eq. (3) from Ref. \cite{YNW}.

\section{Robust Estimator}
Here, we build a robust fixed-interval smoother, corresponding to the optimal smoother above, using the technique from Ref. \cite{MSP} as outlined in \ref{sec:met_rob_smth}.

\subsection{Uncertain Model}
The uncertainty is introduced in the parameter $\lambda$ as follows: \( \lambda \to \lambda - \mu\Delta\lambda,\) where $0 \leq \mu < 1$ determines the level of uncertainty in the model, and $\Delta$ is an uncertain parameter satisfying:
\begin{equation}
||\Delta|| \leq 1,
\end{equation}
which is of the form (\ref{eq:met_smth_unc_1}). Also, the noises $v$ and $w$ are assumed to satisfy the following bound for a suitable constant $d_1>0$:
\begin{equation}
\int_0^\tau (v^2 + w^2) dt \leq d_1,
\end{equation}
which is of the form (\ref{eq:met_smth_unc_2}) with $Q = R = 1$. Moreover, no \emph{a-priori} information exists about the initial condition of the state, and therefore, we choose $X_0 = 0$ in (\ref{eq:met_smth_unc_3}).

Then, the corresponding uncertain system model takes the form:
\begin{equation}
\begin{split}
\dot{\phi} &= -\lambda\phi + B_1\Delta K \phi + B_1 v, \\ 
\theta &= \frac{2|\alpha|}{\sqrt{\overline{R}_{sq}}}\phi + w,\\
\end{split}
\end{equation}
which is of the form (\ref{eq:met_smth_unc_0}). Here $B_1 = \sqrt{\kappa}$ and $K = \mu\lambda/\sqrt{\kappa}$. Also, here $B_2 = G = 0$ in (\ref{eq:met_smth_unc_0}), since there is no known input $u(t)$ in our case. Moreover, here $\Delta_1(t) = \Delta$ and $\Delta_2(t) = 0$ in (\ref{eq:met_smth_unc_0}).

\paragraph{Remark.} As outlined in \ref{sec:met_rob_smth}, the robust fixed-interval smoother takes the form of an ellipse of possible states and the centre of this ellipse is the robust smoother estimate. This robust smoother estimate will be given in terms of two quantities referred to as the forward filter state and the backward filter state, which are defined in the following sections.

\subsection{Forward Filter}
The steady-state forward Riccati equation used in the robust smoother, as obtained from (\ref{eq:met_smth_unc_ric1}), is:
\begin{equation}\label{eq:robust_fwd_riccati}
-2\lambda X + \kappa X^2 + \frac{\mu^2\lambda^2}{\kappa} - \frac{4|\alpha|^2}{\overline{R}_{sq}} = 0.
\end{equation}

The stabilising solution of the above equation for $X$ is:
\begin{equation}\label{eq:robust_fwd_X}
X = \frac{\lambda + \sqrt{\lambda^2 - \mu^2\lambda^2 + \frac{4|\alpha|^2\kappa}{\overline{R}_{sq}}}}{\kappa}.
\end{equation}

Next, the equation (\ref{eq:met_smth_unc_diff1}), that forms part of the robust smoother, for our case yields:
\begin{equation}
\dot{\eta} = -\left(\sqrt{\lambda^2 - \mu^2\lambda^2 + \frac{4|\alpha|^2\kappa}{\overline{R}_{sq}}}\right)\eta + \frac{4|\alpha|^2}{\overline{R}_{sq}}\phi + \frac{2|\alpha|}{\sqrt{\overline{R}_{sq}}}w.
\end{equation}

We then define the quantity, \(\hat{\phi}_f = \eta /X\), which is referred to as the forward filter state.

Thus, the forward robust filter equation is
\begin{equation}\label{eq:robust_fwd_filter}
{\dot{\hat{\phi}}_f = -L\hat{\phi}_f + \frac{4|\alpha|^2\kappa}{\overline{R}_{sq}(\lambda + L)}\phi + \frac{2|\alpha|\kappa}{\sqrt{\overline{R}_{sq}}(\lambda + L)}w,}
\end{equation}
where $L = \sqrt{\lambda^2 - \mu^2\lambda^2 + \frac{4|\alpha|^2\kappa}{\overline{R}_{sq}}}$.

For $\mu=0$, (\ref{eq:robust_fwd_filter}) reduces to (\ref{eq:kalman_fwd_filter}), i.e. the robust forward filter is simply the forward Kalman filter for zero uncertainty level.

\paragraph{Remark.} The quantity $\hat{\phi}_f$ is actually the centre of an ellipse defined by the solution to a robust filtering problem; see Theorem 3.1 in Ref. \cite{MSP}. However, this property will not be used here.

\subsection{Backward Filter}
The steady-state backward Riccati equation for the robust smoother, as obtained from (\ref{eq:met_smth_unc_ric2}), is:
\begin{equation}\label{eq:robust_bwd_riccati}
-2\lambda Y - \kappa Y^2 - \frac{\mu^2\lambda^2}{\kappa} + \frac{4|\alpha|^2}{\overline{R}_{sq}} = 0.
\end{equation}

The stabilising solution of the above equation for $Y$ is:
\begin{equation}\label{eq:robust_bwd_Y}
Y = \frac{-\lambda + \sqrt{\lambda^2 - \mu^2\lambda^2 + \frac{4|\alpha|^2\kappa}{\overline{R}_{sq}}}}{\kappa}.
\end{equation}

Next, the equation (\ref{eq:met_smth_unc_diff2}), that forms part of the robust smoother, in reverse-time yields:
\begin{equation}
\dot{\xi} = -\left(\sqrt{\lambda^2 - \mu^2\lambda^2 + \frac{4|\alpha|^2\kappa}{\overline{R}_{sq}}}\right)\xi + \frac{4|\alpha|^2}{\overline{R}_{sq}}\phi + \frac{2|\alpha|}{\sqrt{\overline{R}_{sq}}}w.
\end{equation}

We then define the quantity, \(\hat{\phi}_b = \xi /Y\), which is referred to as the backward filter state.

Thus, the backward robust filter equation is
\begin{equation}\label{eq:robust_bwd_filter}
{\dot{\hat{\phi}}_b = -L\hat{\phi}_b + \frac{4|\alpha|^2\kappa}{\overline{R}_{sq}(-\lambda + L)}\phi + \frac{2|\alpha|\kappa}{\sqrt{\overline{R}_{sq}}(-\lambda + L)}w,}
\end{equation}
where again $L = \sqrt{\lambda^2 - \mu^2\lambda^2 + \frac{4|\alpha|^2\kappa}{\overline{R}_{sq}}}$.

For $\mu=0$, (\ref{eq:robust_bwd_filter}) reduces to (\ref{eq:kalman_bwd_filter}), i.e. the robust backward filter is the same as the backward Kalman filter for zero uncertainty level.

\paragraph{Remark.} The quantity $\hat{\phi}_b$ is actually the centre of an ellipse defined by the solution to a robust retrodiction (i.e. backward-time filtering) problem; this is similar to the robust (forward) filtering problem considered in Theorem 3.1 in Ref. \cite{MSP}. However, this property will not be used here.

\subsection{Robust Smoother}
The robust smoother estimate is the centre of the ellipse defined in (\ref{eq:met_smth_unc_ellipse}) and is given in terms of $\hat{\phi}_f$ and $\hat{\phi}_b$ according to the following formula:
\begin{equation}\label{eq:robust_smoother}
\hat{\phi} = \frac{X}{X+Y}\hat{\phi}_f + \frac{Y}{X+Y}\hat{\phi}_b.
\end{equation}

\section{Comparison of Estimators}
We shall now compare the mean-square estimation errors of the optimal and robust estimators for the uncertain system.

\subsection{Error Analysis}
Given the forward and backward filter dynamics for the uncertain system, the mean-square errors for $|\Delta|\leq 1$ are computed using the following method employing a Lyapunov equation. Here, we shall illustrate the method for the optimal estimator only. The errors for the robust estimator may be calculated similarly.

\subsubsection{Forward Filter}
The uncertain system, given by
\begin{equation}\label{eq:uncertain_system}
\dot{\phi} = -\lambda\phi+\mu\Delta\lambda\phi+\sqrt{\kappa}v,
\end{equation}
augmented with the forward-time Kalman filter (\ref{eq:kalman_fwd_filter}) may be represented by the state-space model:
\begin{equation}\label{eq:augmented_system}
\dot{\overline{x}} = \overline{A}\, \overline{x} + \overline{B}\, \overline{w},
\end{equation}
where
$\overline{x} = 
\left[ \begin{array}{c}
\phi \\
\hat{\phi}_f
\end{array} \right]$, $\overline{w} = 
\left[ \begin{array}{c}
v \\
w
\end{array} \right]$.

Thus, we have
\[\overline{A} = 
\left[ \begin{array}{cc}
-\lambda+\mu\Delta\lambda & 0 \\
\frac{2|\alpha|K_f}{\sqrt{\overline{R}_{sq}}} & -(\lambda+\frac{2|\alpha|K_f}{\sqrt{\overline{R}_{sq}}})
\end{array} \right],
\quad
\overline{B} = 
\left[ \begin{array}{cc}
\sqrt{\kappa} & 0\\
0 & K_f
\end{array} \right].\]

The steady-state state covariance matrix $C_S$ is obtained by solving the Lyapunov equation:
\begin{equation}\label{eq:lyapunov}
\overline{A}C_S + C_S\overline{A}^T + \overline{B}\, \overline{B}^T = 0,
\end{equation}
where $C_S$ is the symmetric matrix
\begin{equation}\label{eq:fwd_state_cov_mat}
C_S = E[\overline{x}\, \overline{x}^T] =
\left[ \begin{array}{cc}
\Sigma & M_f \\
M_f^T & N_f
\end{array} \right].
\end{equation}

Here, $\Sigma = E[\phi\phi^T]$, $M_f = E[\phi\hat{\phi}_f^T]$, and $N_f = E[\hat{\phi}_f\hat{\phi}_f^T]$.

The estimation error can be written as:
\begin{equation}
e_1 = \phi - \hat{\phi}_f = \left[\begin{array}{cc} 1 & -1 \end{array}\right] \overline{x},
\end{equation}
which is mean zero since all of the quantities determining $e_1$ are mean zero. The error covariance is then given as:
\begin{equation}\label{eq:fwd_err_sym_mat}
\sigma_f^2 = E[e_1e_1^T] = \Sigma - M_f - M_f^T + N_f.
\end{equation} 

\subsubsection{Backward Filter}
When our uncertain model (\ref{eq:uncertain_system}), (\ref{eq:ou_measurement}), which is driven by Gaussian white noise, has reached steady state, the output process will be a stationary Gaussian random process, which is described purely by its auto-correlation function. If we consider this output process in reverse time, this will also be a stationary random process with the same auto-correlation function. This follows from the definition of the auto-correlation function. Hence, the statistics of the reversed time output process are the same as the statistics of the forward time output process. Thus, the reversed time output process can be regarded as being generated by the same (and not time reversed) process (\ref{eq:uncertain_system}) that generated the forward time process \cite{RPH2}.

The augmented system (\ref{eq:augmented_system}) for the backward Kalman filter (\ref{eq:kalman_bwd_filter}) will then have
\[\overline{x} = 
\left[ \begin{array}{c}
\phi \\
\hat{\phi}_b
\end{array} \right],
\qquad
\overline{A} = 
\left[ \begin{array}{cc}
-\lambda+\mu\Delta\lambda & 0 \\
\frac{2|\alpha|K_b}{\sqrt{\overline{R}_{sq}}} & (\lambda-\frac{2|\alpha|K_b}{\sqrt{\overline{R}_{sq}}})
\end{array} \right],
\qquad
\overline{B} = 
\left[ \begin{array}{cc}
\sqrt{\kappa} & 0\\
0 & K_b
\end{array} \right].\]

We then solve (\ref{eq:lyapunov}), with
\begin{equation}\label{eq:bwd_state_cov_mat}
C_S = E[\overline{x}\, \overline{x}^T] =
\left[ \begin{array}{cc}
\Sigma & M_b \\
M_b^T & N_b
\end{array} \right],
\end{equation}
where $\Sigma = E[\phi\phi^T]$, $M_b = E[\phi\hat{\phi}_b^T]$, and $N_b = E[\hat{\phi}_b\hat{\phi}_b^T]$.

The error covariance for the backward filter is then:
\begin{equation}\label{eq:bwd_err_sym_mat}
\sigma_b^2 = E[e_2e_2^T] = \Sigma - M_b - M_b^T + N_b,
\end{equation}
where $e_2 = \phi - \hat{\phi}_b$.

\subsubsection{Smoother Error}
The forward and backward estimates are not independent in general and will have a cross-correlation term as follows \cite{RPH2}:
\begin{equation}\label{eq:crscor_err_sym_mat}
\sigma_{fb}^2 := E[e_1e_2^T] = \Sigma - M_f^T - M_b + \alpha\Sigma\beta,
\end{equation}
where $\alpha = M_f^T\Sigma^{-1}$ and $\beta = \Sigma^{-1}M_b$ \cite{WWS}.

The overall smoother error for the optimal estimator is (see Eq. (25) in Ref. \cite{RPH2}):
\begin{equation}\label{eq:smoothed_error}
{\sigma^2 = \frac{\sigma_f^2\sigma_b^2-(\sigma_{fb}^2)^2}{\sigma_f^2+\sigma_b^2-2\sigma_{fb}^2}.}
\end{equation}

The term $\sigma_{fb}^2$ is zero, and so the forward and backward estimates are independent, in case of the optimal estimator for the exact model, so that (\ref{eq:smoothed_error}) reduces to (\ref{eq:opt_smoother_error}).

However, in the case of the robust estimator, the formula (\ref{eq:smoothed_error}) is replaced by the formula (see Eq. (23) in Ref. \cite{RPH2}):
\begin{equation}\label{eq:smoothed_error_rob}
\sigma^2 = k_1^2\sigma_f^2 + k_2^2\sigma_b^2 + 2k_1k_2\sigma_{fb}^2,
\end{equation}
where $k_1 = \frac{X}{X+Y}$ and $k_2 = \frac{Y}{X+Y}$ from (\ref{eq:robust_smoother}).

\subsection{Comparison of the Errors}
\begin{figure}[!t]
\centering
\includegraphics[width=0.8\textwidth]{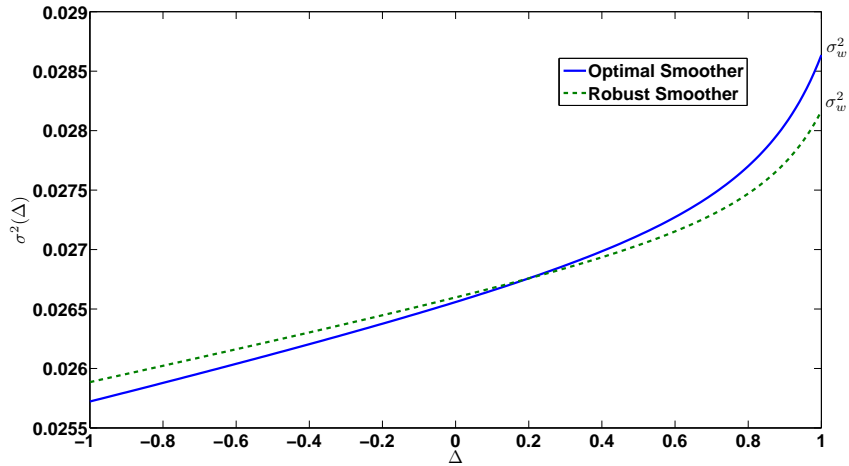}
\caption{Ornstein-Uhlenbeck Noise: Comparison of estimators for uncertainty bound $\mu = 0.8$. Here, $\sigma^2(\Delta)$ is the smoother error covariance of the optimal and robust estimators plotted as a function of the uncertain parameter $|\Delta|\leq 1$.}
\label{fig:robust_vs_optimal}
\end{figure}

The error-covariances of the robust smoother and the optimal smoother for the uncertain system may be computed using the above technique by solving a Lyapunov equation, as a function of $\Delta$. Here, we choose the value of $\mu$ and the other parameters as in Ref. \cite{YNW}, viz. $|\alpha|^2 = 1 \times 10^6$ $\mathrm{s}^{-1}$, $\kappa = 1.9 \times 10^4$ rad/s, $\lambda = 5.9 \times 10^4$ rad/s, $r_m=0.36$ and $r_p=0.59$. Due to the implicit dependence of $\overline{R}_{sq}$ and $\sigma_f^2$ ((\ref{eq:implicit}) and (\ref{eq:fwd_err_sym_mat})), we compute the smoothed mean-square error (\ref{eq:smoothed_error}) and (\ref{eq:smoothed_error_rob}) by running several iterations until $\sigma_f^2$ is obtained with an accuracy of $6$ decimal places in each case. Fig. \ref{fig:robust_vs_optimal} shows a comparison for $\mu = 0.8$, which corresponds to $80\%$ uncertainty in $\lambda$. At $\Delta=0$, where the nominal parameters for the model exactly match those of the system, the optimal smoother performs better than the robust smoother. This is to be expected because the smoother has been optimised for those parameters. However, the robust smoother error is lower than that of the optimal smoother as $\Delta$ approaches $1$. We define $\sigma^2_w(\mu)$ as the worst-case estimation error for each value of $\mu$, i.e. $\sigma^2_w(\mu) = \sigma^2(\Delta=1,\mu)$. So, if our system is not allowed to exceed an error threshold of say $0.0282$ for this level of uncertainty in $\lambda$, our robust estimator guarantees that the error is below this threshold, whereas the optimal estimator breaches it in the worst case.

\begin{figure}[!t]
\centering
\includegraphics[width=0.8\textwidth]{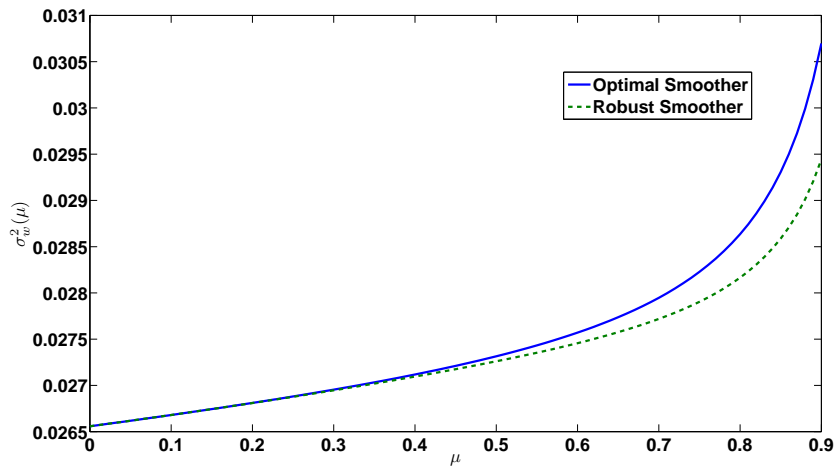}
\caption{Ornstein-Uhlenbeck Noise: Comparison of worst-case error covariances as a function of uncertainty bound $\mu$. Here, $\sigma^2_w(\mu)$ is the worst-case smoother error covariance of the optimal and robust estimators plotted as a function of the uncertainty bound $0\leq\mu\leq 0.9$.}
\label{fig:ou_squeezed_worst}
\end{figure}

Fig. \ref{fig:ou_squeezed_worst} shows the comparison of the worst-case performance of the optimal and the robust estimators for the uncertain system for $0 \leq\mu\leq 0.9$. Clearly, the robust estimator provides with better worst-case performance than the optimal estimator for all levels of uncertainty in $\lambda$. Also, the worst-case robust estimator error is below a desired threshold of say $0.029$ for up to a higher level of uncertainty as compared to the optimal estimator. This is exactly the power of robust design techniques.

\section{Resonant Noise Process}
We now consider a second-order resonant noise process, typically produced by a piezo-electric transducer (PZT) driven by an input white noise. Such a resonant process is more complicated than the simplistic OU noise process considered before and better resembles the kind of noises that in practice corrupt the signal. The simplified transfer function of a typical PZT is (see the supplementary material of Ref. \cite{KI}) as follows:
\begin{equation}\label{eq:pzt_tf}
{G(s) := \frac{\phi(s)}{v(s)} = \frac{\kappa}{s^2+2\zeta\omega_r s+\omega_r^2},}
\end{equation}
where $\kappa$ is the gain, $\zeta$ is the damping factor, $\omega_r$ is the resonant frequency (rad/s), $v$ is a zero-mean white Gaussian noise with unity amplitude and $\phi$ is the PZT output that modulates the phase to be estimated.

\subsection{System Model (Exact)}
A state-space realization of the transfer function (\ref{eq:pzt_tf}) is:
\begin{equation}\label{eq:process_eqn}
\dot{x} = Ax + Bv,
\end{equation}
where
\[x := \left[\begin{array}{c}
\phi\\
\dot{\phi}
\end{array}\right], \quad
A := \left[\begin{array}{cc}
0 & 1\\
-\omega_r^2 & -2\zeta\omega_r
\end{array}\right], \quad
B := \left[\begin{array}{c}
0\\
\kappa
\end{array}\right].\]

Eq. (\ref{eq:process_eqn}) constitutes our process model, whereas the measurement remains the same as (\ref{eq:ou_measurement}). Thus, our measurement equation is
\begin{equation}\label{eq:measurement_eqn}
\theta = Cx+w,
\end{equation}
where $C := \left[\begin{array}{cc} 2|\alpha|/\sqrt{\overline{R}_{sq}} & 0 \end{array}\right]$.

In this paper, we choose parameter values which are particularly suited to the illustration of our key robustness results and yet represent a possible physical situation, viz. $\kappa = 9 \times 10^4$, $\zeta = 0.1$ and $\omega_r = 6.283 \times 10^3$ rad/s and $|\alpha|^2 = 25 \times 10^4$ $\mathrm{s}^{-1}$.

\subsection{Uncertain Model}
We introduce uncertainty in $A$ as follows:
\begin{equation}\label{eq:sqz_unc_model}
A \to A + \left[\begin{array}{cc}
0 & 0\\
-\mu\delta\omega_r^2 & 0
\end{array}\right],
\end{equation}
where uncertainty is introduced in the resonant frequency $\omega_r$ through $\delta$. Although uncertainty in $\omega_r$ would affect both entries in the second row of the above matrix, the most significant effect will be in the $\omega_r^2$ term. Indeed, since we have a resonant system and $\zeta \ll \omega_r$, the uncertainty in $-2\zeta\omega_r$ term can be neglected for simplicity and to give a less conservative estimator.

Here, $\Delta := \delta $ is an uncertain parameter satisfying $||\Delta|| \leq 1$ which implies $\delta^2 \leq 1$. Moreover, $\mu \in [0,1)$ determines the level of uncertainty. From (\ref{eq:met_smth_unc_0}), the uncertain model here is:
\begin{equation}\label{eq:uncertain_model}
\begin{split}
\textsf{\small Process model:} \ \ \dot{x} &= (A+B\Delta K)x + Bv, \\
\textsf{\small Measurement model:} \ \ \theta &= Cx + w,
\end{split}
\end{equation}
where 
$K:=\left[\begin{array}{cc}
-\frac{\mu\omega_r^2}{\kappa} & 0
\end{array}\right]$.

\subsection{Comparison of the Estimators}
The optimal and robust estimators can be constructed for the resonant noise case using the same method employed in the OU noise case before. The mean-square errors in estimation of $\phi$ may be computed using the error-analysis technique discussed before for both the optimal smoother and robust fixed-interval smoother as a function of the uncertain parameter $\delta$. Note that $P_f$, $K_f$ in (\ref{eq:fwd_err_cov_kalman}),(\ref{eq:kalman_fwd_filter}), $P_b$, $K_b$ in (\ref{eq:bwd_err_cov_kalman}),(\ref{eq:kalman_bwd_filter}), $X$ in (\ref{eq:robust_fwd_X}), and $Y$ in (\ref{eq:robust_bwd_Y}) are $2\times2$ matrices, and not scalars, in this resonant noise case. Thus, the expressions in (\ref{eq:fwd_err_sym_mat}), (\ref{eq:bwd_err_sym_mat}) and (\ref{eq:crscor_err_sym_mat}) yield $2 \times 2$ matrices, and not scalars, in this resonant noise case for both the optimal and robust smoothers, but the values that we use to compute the effective smoother error (\ref{eq:smoothed_error}) and (\ref{eq:smoothed_error_rob}) here are the $(1,1)$ entries in these matrices, since we are interested in the estimation errors in $\phi$ and not $\dot{\phi}$ from $x$ in (\ref{eq:uncertain_model}). These values can be used to generate a plot of the errors versus $\delta$ for a given value of $\mu$ to compare the performance of the robust smoother and the optimal smoother for the uncertain system. Here, we used the nominal parameter values, and $r_m=0.48$ and $r_p=1.11$ to have an optimal squeezing level, for which the estimation error is the minimum for the exact model \cite{YNW}. Again, due to the implicit dependence of $\overline{R}_{sq}$ and $\sigma_f^2$, we compute the smoothed mean-square error (\ref{eq:smoothed_error}) and (\ref{eq:smoothed_error_rob}) by running several iterations until $\sigma_f^2$ is obtained with an accuracy of $6$ decimal places in each case.

\begin{figure}[!t]
\centering
\includegraphics[width=0.8\textwidth]{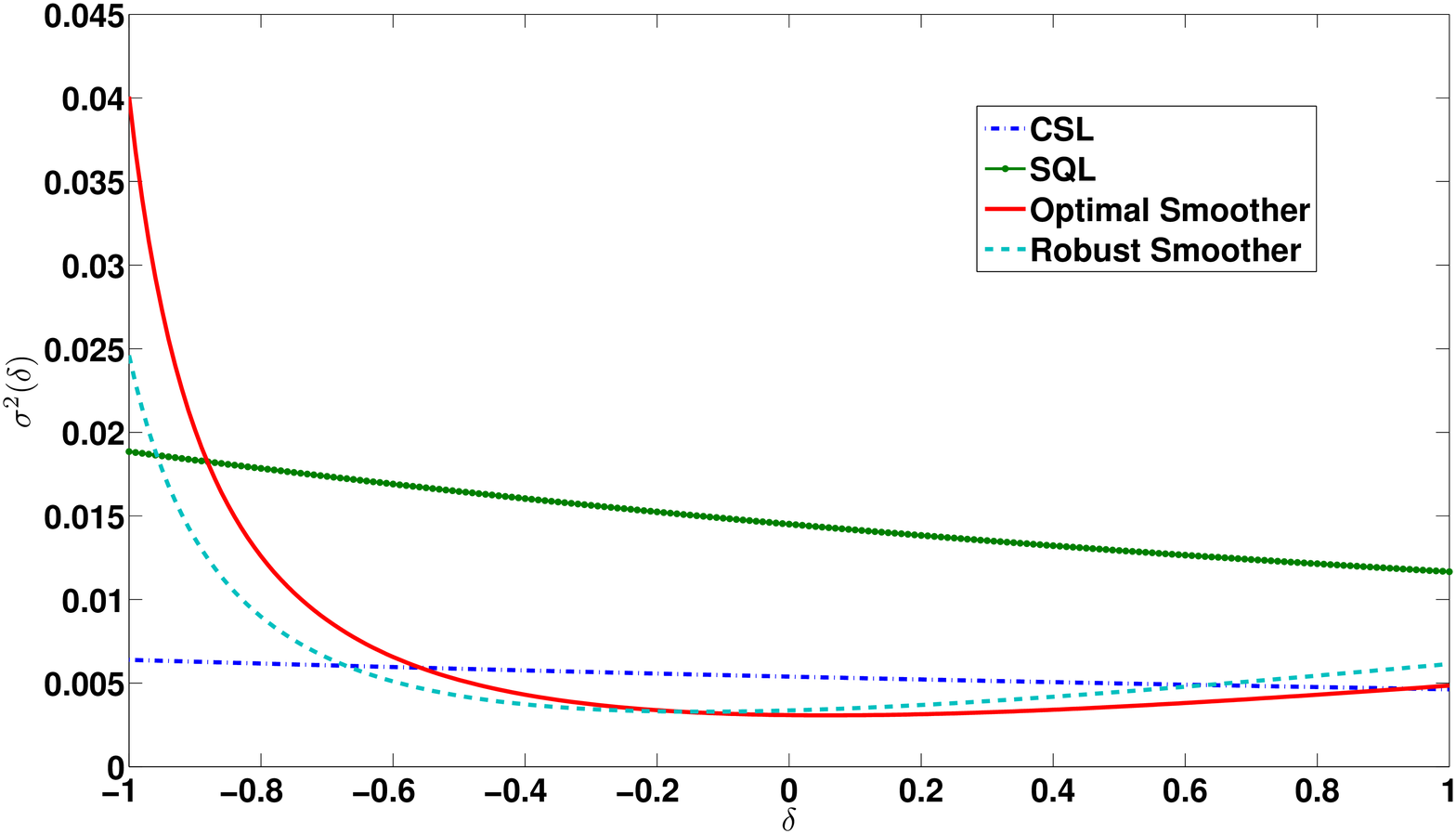}
\caption{Resonant Noise: Comparison of the smoothers for uncertainty level $\mu=0.8$. Here, $\sigma^2(\delta)$ is the smoother error covariance of the optimal and robust estimators plotted as a function of the uncertain parameter $|\delta|\leq 1$. Moreover, CSL = Coherent State Limit and SQL = Standard Quantum Limit.}
\label{fig:res_sqz_delta}
\end{figure}

It is also insightful to include in the graph the coherent state limit (CSL), which is the minimum theoretical error reachable with a coherent beam \cite{YNW}. The CSL value is obtained by designing a different optimal smoother for each value of the uncertain parameter in (\ref{eq:uncertain_model}) (note $r_m$, $r_p = 0$ and $\overline{R}_{sq} = 1$ for coherent beam), and is given by $\sigma^2 = P_s(1,1)$ from (\ref{eq:opt_smoother_error}). We as well include the standard quantum limit (SQL), which is the minimum phase estimation error that can be obtained with coherent beam using perfect heterodyne technique. The method to compute the SQL for our resonant noise model is given in \ref{sec:sql_resonant}. The SQL value is obtained for our plots for each value of the uncertain parameter in the process model in (\ref{eq:uncertain_model}), with the measurement model in (\ref{eq:dual_hd_meas}). Fig. \ref{fig:res_sqz_delta} shows the plot for $\mu=0.8$.

One can see that the optimal smoother behaves better than the robust smoother when $\delta=0$, as expected. However, in the worst-case scenario, i.e. as $\delta$ approaches $-1$, the performance of the robust smoother is superior to that of the optimal smoother. Nonetheless, we trade off the best-case performance in achieving it. It is also of relevance that the robust estimator beats the SQL over a larger part of the uncertainty window than the optimal estimator, although this is not the case with respect to the CSL for this value of $\mu$.

\begin{figure}[!t]
\centering
\includegraphics[width=0.8\textwidth]{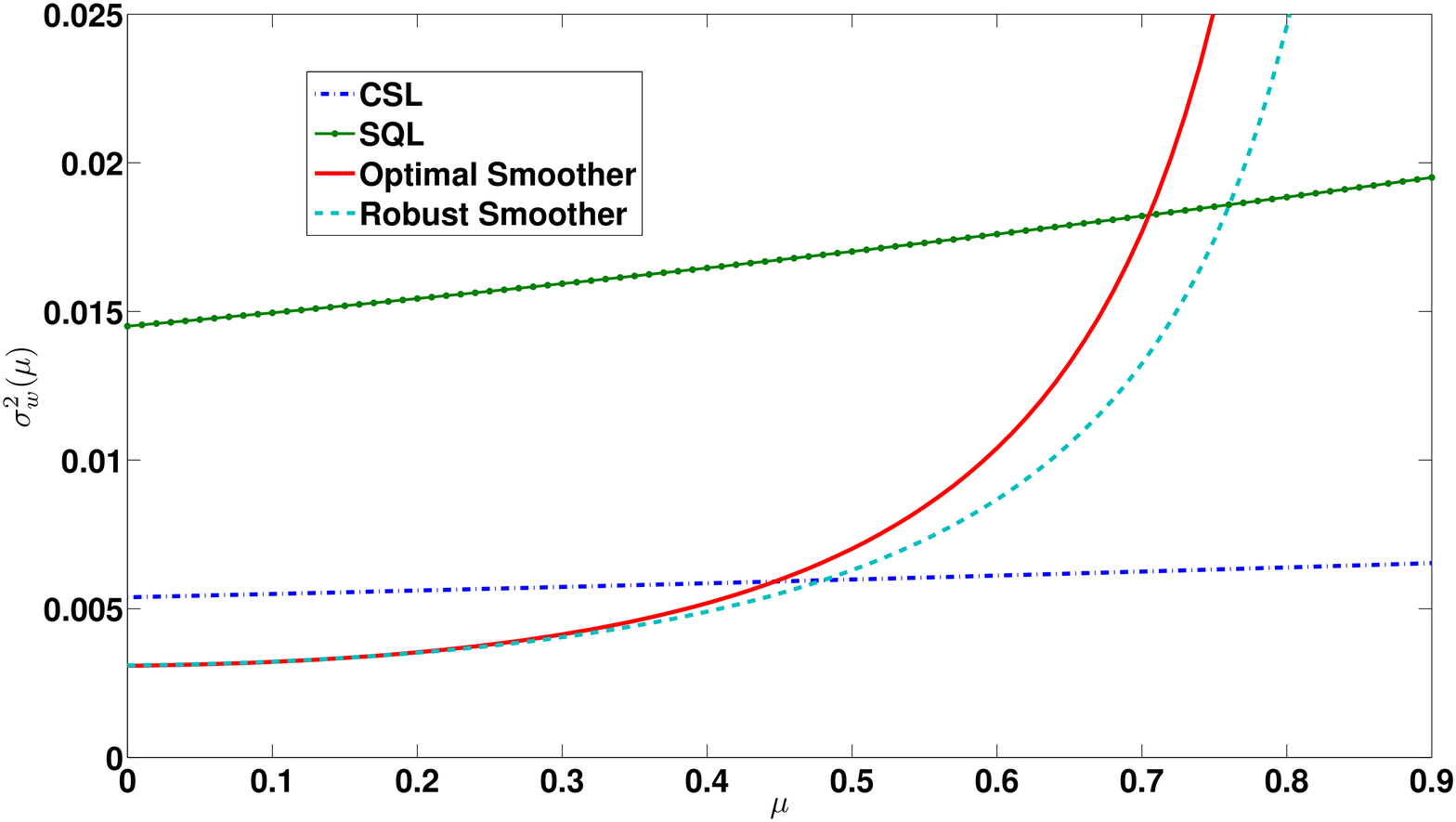}
\caption{Resonant Noise: Comparison of worst-case error covariances as a function of uncertainty level $\mu$. Here, $\sigma^2_w(\mu)$ is the worst-case smoother error covariance of the optimal and robust estimators plotted as a function of the uncertainty level $0\leq\mu\leq 0.9$. Moreover, CSL = Coherent State Limit and SQL = Standard Quantum Limit.}
\label{fig:res_sqz_worst}
\end{figure}

Fig. \ref{fig:res_sqz_worst} depicts the worst-case performance of the estimators for $0\leq\mu\leq 0.9$. Clearly, the robust estimator outperforms the optimal estimator in the worst-case for all levels of $\mu$. Also shown in the plot are the SQL and the CSL. If the SQL or the CSL is considered as the allowed threshold for the estimation error, our robust estimator provides guaranteed worst-case performance below this threshold for up to a larger uncertainty level when compared to the optimal estimator.

\begin{figure}[!b]
\centering
\includegraphics[width=0.8\textwidth]{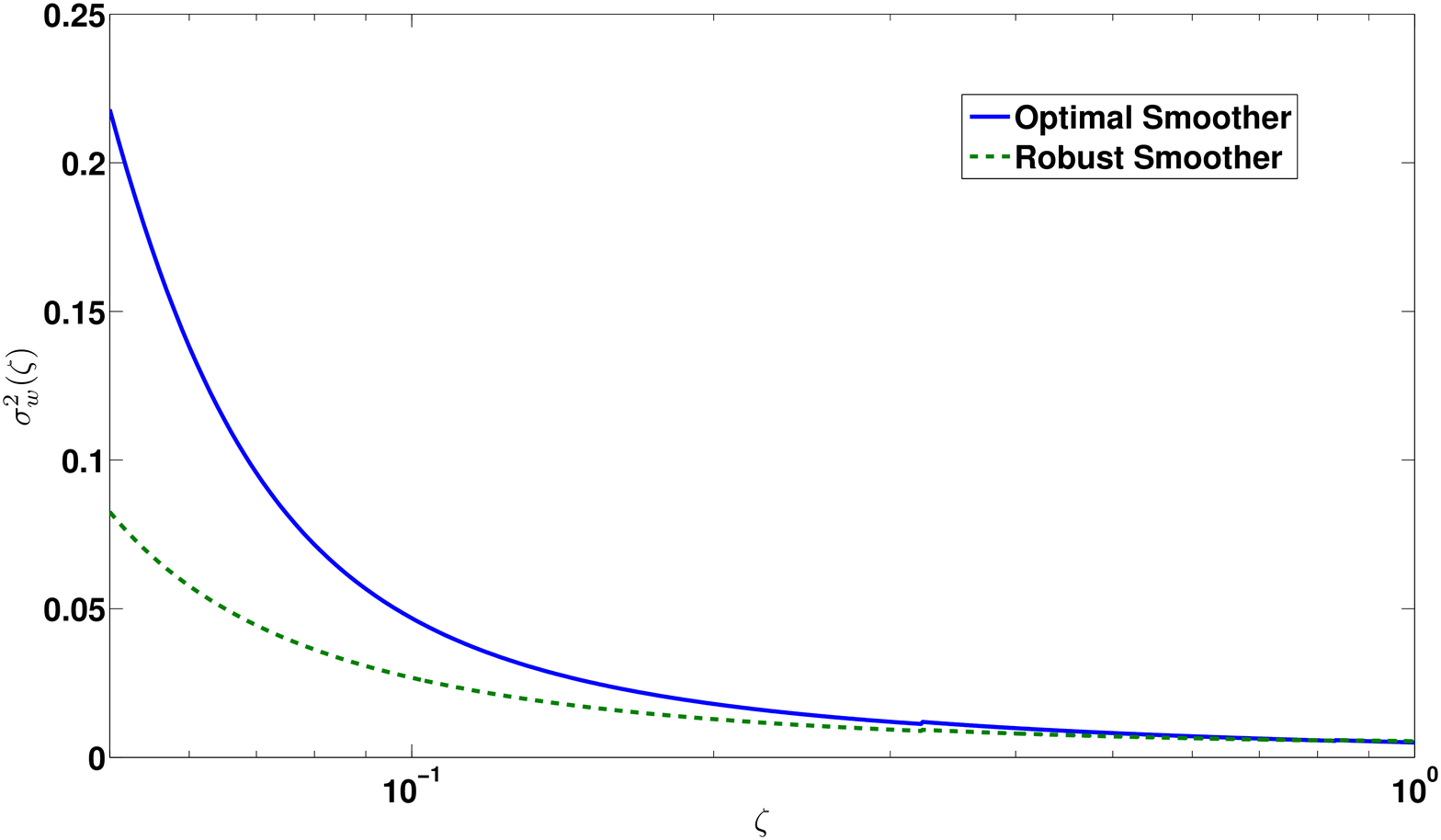}
\caption{Resonant Noise: Comparison of worst-case error covariances as a function of damping factor $\zeta$. Here, $\sigma^2_w(\zeta)$ is the worst-case smoother error covariance of the optimal and robust estimators plotted as a function of the damping factor $\zeta$ varying from $0.05$ to $1$.}
\label{fig:res_sqz_worst_zeta}
\end{figure}

Moreover, the improvement with the robust smoother over the optimal smoother is better with the resonant noise process considered here as compared to that with OU noise process considered before. For example, while the worst-case improvement for $80\%$ uncertainty in the OU noise case was $\sim 0.08$ dB, that in this resonant noise case is $\sim 2.13$ dB. Indeed, OU noise is the output of a non-resonant low-pass filter (LPF), driven by white noise. Any uncertainty in the corner frequency of the LPF, represented by $\lambda$ here, would not change the magnitude of the phase noise as much as an equivalent amount of uncertainty in the resonant frequency $\omega_r$ for the resonant noise model. In fact, the relative performance of our robust estimator grows as the noise process becomes more resonant. This is shown in Fig. \ref{fig:res_sqz_worst_zeta}, which plots the worst-case errors of the optimal and robust smoothers as a function of the damping factor $\zeta$, that determines the degree of resonance; i.e. the lower damping factor, the more resonant the process is. Here, at each value of $\zeta$, the squeezing level has been optimized to yield the minimum error for the exact model.

\begin{figure}[!t]
\centering
\includegraphics[width=0.8\textwidth]{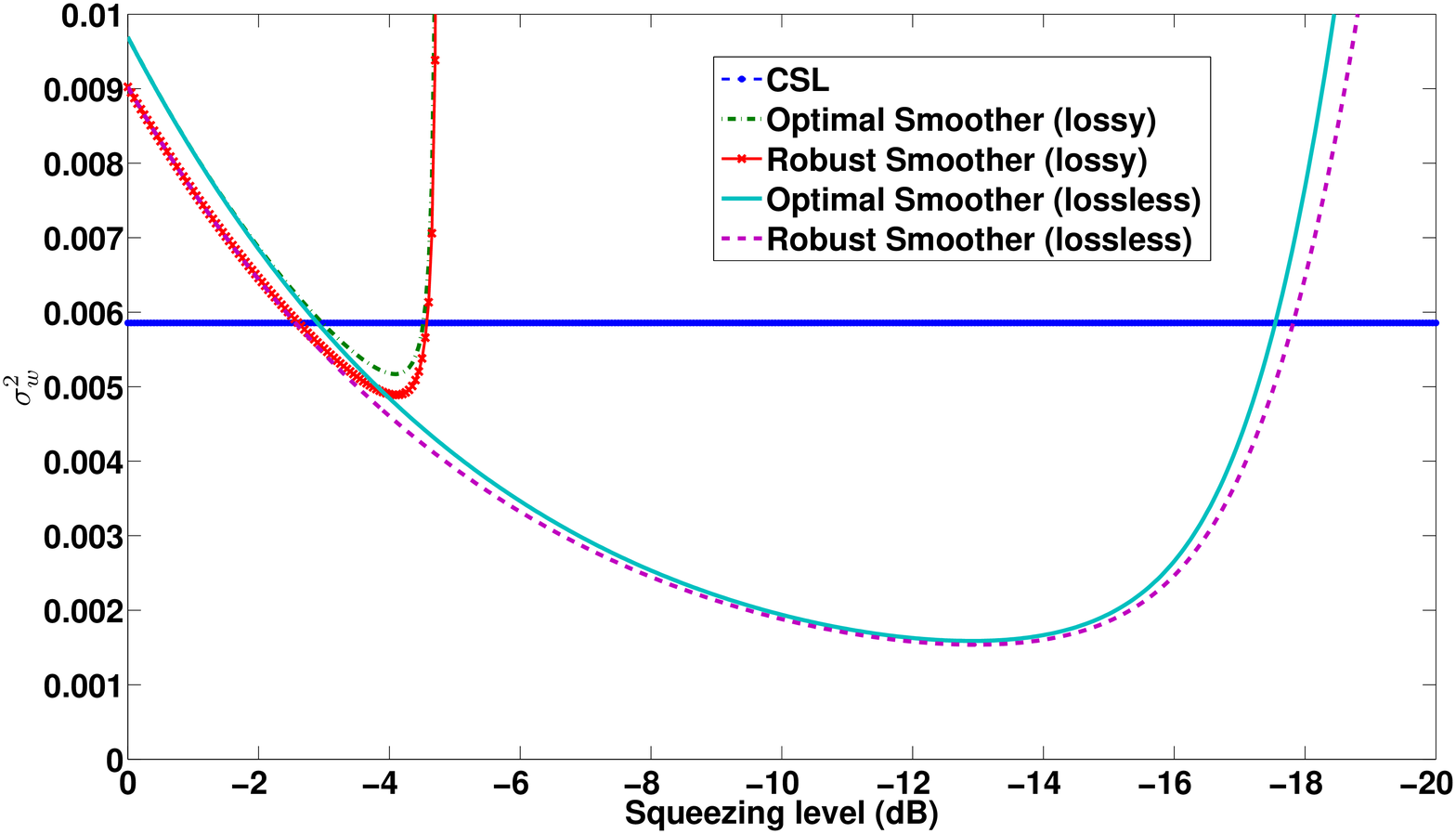}
\caption{Resonant Noise: Comparison of worst-case error covariances as a function of squeezing level. Here, $\sigma^2_w$ is the worst-case smoother error covariance of the optimal and robust estimators plotted as a function of the squeezing level varying from 0 to -20 dB. Moreover, CSL = Coherent State Limit.}
\label{fig:res_sqz_worst_sqzlvl}
\end{figure}

Fig. \ref{fig:res_sqz_worst_sqzlvl} shows a plot of the worst-case errors of the smoothers as functions of the squeezing level. Here, we have plotted the errors for pure lossless squeezed beams (overall loss $l_{sq}=0$) and practical impure lossy squeezed beams with $l_{sq}=0.33$ (see the supplementary material for Ref. \cite{YNW}), and a fixed uncertainty level of $\mu=0.4$. Clearly, our robust estimator not only beats the CSL over a wider range of squeezing levels (for both pure and impure squeezing cases), but also can sustain higher levels of squeezing than the optimal estimator can, before the errors rapidly increase due to excessive anti-squeezing noise. Moreover, our estimator is more robust relative to the optimal estimator for practical squeezed beams than for ideal pure squeezed beams. This is due to the larger worst-case performance benefit obtained in the practical case than in the ideal case at the optimal squeezing levels. While the robust estimation worst-case error is $\sim 0.15$ dB lower than the optimal estimation worst-case error at the optimal squeezing level of $-12.9$ dB for the lossless case, the robust estimation worst-case performance is $\sim 0.26$ dB better than the optimal estimation worst-case performance at the optimal squeezing level of $-4.1$ dB for the lossy case. That is, our estimator, which was designed to be robust to uncertainty in the resonant frequency $\omega_r$, is also robust, relative to the optimal estimator, against the overall loss $l_{sq}$, arising from imperfect detectors, the optical parametric oscillator (OPO) and modulators.

\begin{figure}[!t]
\centering
\includegraphics[width=0.8\textwidth]{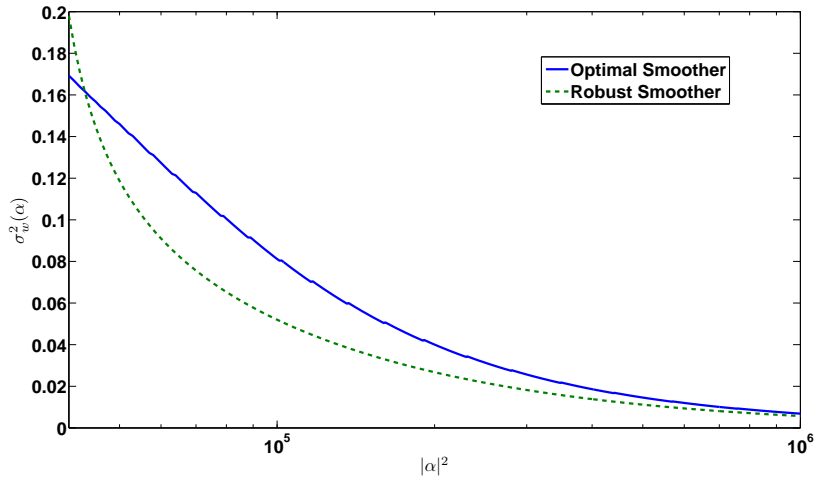}
\caption{Resonant Noise: Comparison of worst-case error covariances as a function of photon flux. Here, $\sigma^2_w(\alpha)$ is the worst-case smoother error covariance of the optimal and robust estimators plotted as a function of the photon flux varying from $4 \times 10^4$ to $10^6$ $\mathrm{s}^{-1}$.}
\label{fig:res_sqz_worst_flux}
\end{figure}

Finally, we plot the worst-case errors of the estimators against the photon flux $|\alpha|^2$ in Fig. \ref{fig:res_sqz_worst_flux}. Here, at each value of $|\alpha|^2$, the squeezing level has been optimized to yield the least worst-case robust estimation error. One can choose to optimize the squeezing level on a different basis as well. Interestingly, not only do the two errors scale differently with the photon flux, but also note there exists an optimum photon flux, and therefore an optimum photon number, for which the robust estimator provides the best worst-case performance compared to the optimal estimator. This is quite significant, given how important the achievable precision for available finite quantum resources is in practice.

\section{Conclusion}
This work considered robust quantum phase estimation with explicitly modelled uncertainty introduced in the underlying system in a systematic state-space setting within the modern control theory paradigm. In particular, we constructed a robust fixed-interval smoother for continuous phase estimation of a squeezed state of light with uncertainty considered in the phase noise. We illustrated that our robust estimator provides guaranteed worst-case performance as desired. We showed that the worst-case performance of our robust estimator with respect to the optimal estimator improves with greater resonance in the phase noise. Moreover, we found that robustness is more useful for practical lossy squeezed beams, when compared to pure squeezed beams, ideally limited by Heisenberg's uncertainty principle. In addition, we saw that there is an optimal photon number for which the performance of the robust estimator relative to the optimal estimator is the best. These results demonstrate the significant impact that the rich theory of classical robust estimation can have on improving quantum parameter estimation. They can pave the way for tackling practical challenges owing to unavoidable parametric uncertainties facing quantum parameter estimation.

\section*{Acknowledgments}
This work was supported by the Australian Research Council. The first author would like to thank Dr. Hongbin Song, Dr. Obaid Ur Rehman, Trevor Wheatley, Prof. Howard Wiseman and Dr. Dominic Berry for useful discussion and feedback related to this work.

\appendix

\section{Kalman Filtering and Optimal Smoothing Theory}\label{sec:met_opt_smth}
We first outline here the continuous-time formulation of the Kalman filter, called the Kalman-Bucy filter. Then we outline the optimal two-filter smoothing theory, as discussed in Ref. \cite{RGB}, but using our notation.

\subsection{Kalman Filter}
The process and measurement models are assumed to be of the form:
\begin{equation}\label{eq:met_system_model}
\begin{split}
\textsf{\small Process model:} \ \ \dot{x} &= Ax + Bv,\\
\textsf{\small Measurement model:} \ \ y &= Cx + Dw,
\end{split}
\end{equation}
where
\begin{equation}\label{eq:met_kalman_covs}
\begin{split}
E[v(t)v^T(r)] &= N\delta(t-r),\\
E[w(t)w^T(r)] &= S\delta(t-r),\\
E[v(t)w^T(r)] &= 0.
\end{split}
\end{equation}
Here, the noises $v(t)$ and $w(t)$ are assumed to be vector white-noise processes with zero cross-correlation. Also, $x(t)$ is the state of the process to be estimated and $y(t)$ is the measurement output. Note that the matrices $A$, $B$, $C$ and $D$ may be time-varying.

Then, the error-covariance matrix $P$ of the Kalman filter is the stabilizing solution of the following matrix differential Riccati equation:
\begin{equation}\label{eq:met_kalman_riccati}
\dot{P} = AP + PA^T - PC^T(DSD^T)^{-1}CP + BNB^T, \qquad P(0) = P_0,
\end{equation}
where $P_0$ is the initial error-covariance.

In the steady-state case, the Riccati equation to be solved to construct the Kalman filter is the following \emph{algebraic} Riccati equation:
\begin{equation}\label{eq:met_kalman_alg_riccati}
AP + PA^T - PC^T(DSD^T)^{-1}CP + BNB^T = 0.
\end{equation}

Note that the above Riccati equation is quadratic in $P$.

The gain of the Kalman filter, called the Kalman gain, is then given as:
\begin{equation}\label{eq:met_kalman_gain}
K_g = PC^T(DSD^T)^{-1}.
\end{equation}

The continuous Kalman filter equation is given as:
\begin{equation}\label{eq:met_kalman_filter}
\dot{\hat{x}} = A\hat{x}+K_g(y-C\hat{x}), \qquad \hat{x}(0) = x_0,
\end{equation}
where $\hat{x}$ is the desired estimate of the state $x$ and $x_0$ is the initial state estimate.

\subsection{Optimal Smoother}
The two-filter smoother, as its name suggests, consists of two different filters, one forward-time, and one backward-time, whose estimates are combined to yield a final smoothed estimate \cite{RGB}.

Let us assume that data is available over a fixed time-interval $[0,\tau]$, and we desire an optimal smoothed estimate $\hat{x}_s(t')$ at a point $0<t'<\tau$. The forward-time variables will be denoted with subscripts, such as $\hat{x}_f$, and backward-time variables with subscripts, such as $\hat{x}_b$. The process and measurement models for the forward-time filter would be:
\begin{equation}
\begin{split}
\dot{x} &= Ax+Bv,\\
y &= Cx+Dw,
\end{split}
\end{equation}
where (\ref{eq:met_kalman_covs}) holds. The time variable in this case is $t$, running forward in time.

Then, the steady-state Riccati equation to be solved for the forward filter is:
\begin{equation}
AP_f+P_fA^T-P_fC^T(DSD^T)^{-1}CP_f+BNB^T=0.
\end{equation}

Also the filter equation is given as:
\begin{equation}
\dot{\hat{x}}_f = A\hat{x}_f+K_f(y-C\hat{x}_f), \qquad \hat{x}_f(0) = 0.
\end{equation}

Here, the forward Kalman gain is:
\begin{equation}
K_f = P_fC^T(DSD^T)^{-1}.
\end{equation}

For the backward filter, it is convenient to define a new running time variable $q$ that proceeds backward in time. Note that $q = 0$ corresponds to $t=\tau$. The backward process model is then obtained by replacing the time derivative in the above process model with $-d/dq$:
\begin{equation}
\frac{dx}{dq} = -Ax-Bv.
\end{equation}

The Riccati and filter equations may then be obtained by replacing $A$ and $B$ in the corresponding equations for the forward filter with $-A$ and $-B$, respectively.

The backward filter steady-state Riccati equation is:
\begin{equation}
-AP_b-P_bA^T-P_bC^T(DSD^T)^{-1}CP_b+BNB^T=0.
\end{equation}

The backward filter equation is:
\begin{equation}
\frac{d\hat{x}_b}{dq} = -A\hat{x}_b + K_b(y-C\hat{x}_b).
\end{equation}

Here, the backward Kalman gain is:
\begin{equation}
K_b = P_bC^T(DSD^T)^{-1}.
\end{equation}

The smoothing error-covariance is then computed as:
\begin{equation}
P_s(t') = (P_f^{-1}+P_b^{-1})^{-1}.
\end{equation}

The equation for the smoothed estimate is then:
\begin{equation}
\hat{x}_s(t') = P_s(t')[P_f^{-1}\hat{x}_f+P_b^{-1}\hat{x}_b].
\end{equation}

\section{Robust Fixed-Interval Smoothing Theory}\label{sec:met_rob_smth}
We outline here the robust fixed-interval smoothing theory from Ref. \cite{MSP}, but using our notation. Consider an uncertain system described by the state equations
\begin{equation}\label{eq:met_smth_unc_0}
\begin{split}
\dot{x}(t) &= [A + B_1\Delta_1(t)K]x(t) + B_1v(t) + [B_2+B_1\Delta_1(t)G]u(t),\\
y(t) &= [C + \Delta_2(t)K]x(t) + w(t) + \Delta_2(t)Gu(t),
\end{split}
\end{equation}
where $x(t)$ is the state, $y(t)$ is the measured output, $u(t)$ is a known input, $v(t)$ and $w(t)$ are noises. $A, B_1, B_2, K, G$ and $C$ are matrices. Furthermore, $\Delta_1(t)$ and $\Delta_2(t)$ are uncertainty matrices satisfying
\begin{equation}\label{eq:met_smth_unc_1}
\left|\left|\left[\begin{array}{cc} \Delta_1(t)^TQ^{\frac{1}{2}} & \Delta_2(t)^TR^{\frac{1}{2}} \end{array}\right]\right|\right| \leq 1
\end{equation}
for all $t$, where $Q = Q^T > 0$ and $R = R^T > 0$ are weighting matrices. Then, for a given finite time-interval $[0, \tau]$ and a given suitable constant $d_1 > 0$, we require the noises to satisfy the inequality
\begin{equation}\label{eq:met_smth_unc_2}
\int_0^\tau \left(v(t)^TQv(t) + w(t)^TRw(t)\right)dt \leq d_1.
\end{equation}
Let $X_0 = X_0^T > 0$ be a given matrix, $x_0$ be a given real vector, $d_2 > 0$ be a given suitable constant. Then, we assume the initial conditions $x(0)$ satisfy the inequality
\begin{equation}\label{eq:met_smth_unc_3}
(x(0)-x_0)^TX_0(x(0)-x_0) \leq d_2.
\end{equation}
This uncertain system is a special case of the uncertain system considered in Eq. (3.19) of Ref. \cite{MSP} of the following form:
\begin{equation}\label{eq:met_smth_unc_4}
\begin{split}
\dot{x}(t) &= Ax(t) + B_1\tilde{v}(t) + B_2u(t),\\
y(t) &= Cx(t) + \tilde{w}(t),\\
z(t) &= Kx(t) + Gu(t),
\end{split}
\end{equation}
where the output $z(t)$ defines the structure of the uncertainty in the uncertain system model, and the quantities $\tilde{v}(t)$ and $\tilde{w}(t)$ are given by
\begin{equation}\label{eq:met_smth_unc_5}
\begin{split}
\tilde{v}(t) &= \Delta_1(t)z(t) + v(t),\\
\tilde{w}(t) &= \Delta_2(t)z(t) + w(t).
\end{split}
\end{equation}

Then, (\ref{eq:met_smth_unc_0}) is obtained by substituting (\ref{eq:met_smth_unc_5}) into (\ref{eq:met_smth_unc_4}).

Let us consider weighting matrices $\tilde{Q} = \tilde{Q}^T > 0$ and $\tilde{R} = \tilde{R}^T > 0$. Then, using (\ref{eq:met_smth_unc_5}), we have:
\begin{equation}\label{eq:met_smth_unc_6}
\begin{split}
\tilde{v}(t)^T\tilde{Q}\tilde{v}(t) &= z(t)^T\Delta_1(t)^T\tilde{Q}\Delta_1(t)z(t) + z(t)^T\Delta_1(t)^T\tilde{Q}v(t)\\
&+ v(t)^T\tilde{Q}\Delta_1(t)z(t) + v(t)^T\tilde{Q}v(t),\\
\tilde{w}(t)^T\tilde{R}\tilde{w}(t) &= z(t)^T\Delta_2(t)^T\tilde{R}\Delta_2(t)z(t) + z(t)^T\Delta_2(t)^T\tilde{R}w(t)\\
&+ w(t)^T\tilde{R}\Delta_2(t)z(t) + w(t)^T\tilde{R}w(t).
\end{split}
\end{equation}

Also, for a given constant $\epsilon > 0$, the following holds, since $\tilde{Q}>0$:
\begin{equation}
\left(\epsilon\Delta_1(t)z(t)-\frac{1}{\epsilon}v(t)\right)^T\tilde{Q} \left(\epsilon\Delta_1(t)z(t)-\frac{1}{\epsilon}v(t)\right) \geq 0.
\end{equation}

This implies
\begin{equation}\label{eq:met_smth_unc_7}
\begin{split}
z(t)^T\Delta_1(t)^T\tilde{Q}v(t)+v(t)^T\tilde{Q}\Delta_1(t)z(t) &\leq \epsilon^2z(t)^T\Delta_1(t)^T\tilde{Q}\Delta_1(t)z(t)\\
&+ \frac{1}{\epsilon^2}v(t)^T\tilde{Q}v(t).
\end{split}
\end{equation}

Similarly, since $\tilde{R}>0$, we have
\begin{equation}
\left(\epsilon\Delta_2(t)z(t)-\frac{1}{\epsilon}w(t)\right)^T\tilde{R} \left(\epsilon\Delta_2(t)z(t)-\frac{1}{\epsilon}w(t)\right) \geq 0.
\end{equation}

This implies
\begin{equation}\label{eq:met_smth_unc_8}
\begin{split}
z(t)^T\Delta_2(t)^T\tilde{R}w(t)+w(t)^T\tilde{R}\Delta_2(t)z(t) &\leq \epsilon^2z(t)^T\Delta_2(t)^T\tilde{R}\Delta_2(t)z(t)\\
&+ \frac{1}{\epsilon^2}w(t)^T\tilde{R}w(t).
\end{split}
\end{equation}

Then, it follows from (\ref{eq:met_smth_unc_6}), (\ref{eq:met_smth_unc_7}), and (\ref{eq:met_smth_unc_8}) that:
\begin{equation}\label{eq:met_smth_unc_9}
\begin{split}
\tilde{v}(t)^T\tilde{Q}\tilde{v}(t) + \tilde{w}(t)^T\tilde{R}\tilde{w}(t) &\leq (1+\epsilon^2)z(t)^T\Delta_1(t)^T\tilde{Q}\Delta_1(t)z(t)\\
&+ (1+\epsilon^2)z(t)^T\Delta_2(t)^T\tilde{R}\Delta_2(t)z(t)\\
&+ \left(1+\frac{1}{\epsilon^2}\right)v(t)^T\tilde{Q}v(t) + \left(1+\frac{1}{\epsilon^2}\right)w(t)^T\tilde{R}w(t).
\end{split}
\end{equation}

Now, we let
\begin{equation}\label{eq:met_smth_unc_10}
\begin{split}
\tilde{Q} = \frac{1}{1+\epsilon^2}Q > 0,\\
\tilde{R} = \frac{1}{1+\epsilon^2}R > 0.
\end{split}
\end{equation}

Since $\epsilon>0$ can be chosen to be arbitrarily small, then $\tilde{Q}$ and $\tilde{R}$ will be arbitrarily close to $Q$ and $R$, respectively.

Thus, using (\ref{eq:met_smth_unc_10}) and (\ref{eq:met_smth_unc_1}) in (\ref{eq:met_smth_unc_9}) and integrating, we get
\begin{equation}\label{eq:met_smth_unc_11}
\begin{split}
\int_0^\tau (\tilde{v}(t)^T\tilde{Q}\tilde{v}(t) + \tilde{w}(t)^T\tilde{R}\tilde{w}(t)) dt &\leq \int_0^\tau z^Tz dt\\
&+ \left(\frac{1+\frac{1}{\epsilon^2}}{1+\epsilon^2}\right) \int_0^\tau (v(t)^TQv(t) + w(t)^TRw(t)) dt\\
&\leq \int_0^\tau ||z(t)||^2 dt + \left(\frac{1+\frac{1}{\epsilon^2}}{1+\epsilon^2}\right) d_1,
\end{split}
\end{equation}
where we have used (\ref{eq:met_smth_unc_2}).

We now let
\begin{equation}\label{eq:met_smth_unc_12}
d = \left(\frac{1+\frac{1}{\epsilon^2}}{1+\epsilon^2}\right) d_1 + d_2 > 0
\end{equation}

Then, it follows from (\ref{eq:met_smth_unc_3}), (\ref{eq:met_smth_unc_11}) and (\ref{eq:met_smth_unc_12}) that the following integral quadratic constraint (IQC) is satisfied by the uncertainty in the system (\ref{eq:met_smth_unc_4}):
\begin{equation}\label{eq:met_smth_unc_iqc}
(x(0)-x_0)^TX_0(x(0)-x_0) + \int_0^\tau (\tilde{v}(t)^T\tilde{Q}\tilde{v}(t)
+\tilde{w}(t)^T\tilde{R}\tilde{w}(t))dt \leq d + \int_0^\tau ||z(t)||^2 dt,
\end{equation}
which corresponds to Eq. (3.20) in Ref. \cite{MSP}.

As mentioned above, $\tilde{Q}$ and $\tilde{R}$ can be chosen to be arbitrarily close to $Q$ and $R$, respectively. For simplicity, in the sequel, we will take $\tilde{Q}=Q$ and $\tilde{R}=R$.

A steady-state solution to the robust fixed-interval smoothing problem for this uncertain system involves the algebraic Riccati equations:
\begin{align}\label{eq:met_smth_unc_ric1}
XA + A^TX + XB_1Q^{-1}B_1^TX+K^TK - C^TRC&=0,\\
YA + A^TY - YB_1Q^{-1}B_1^TY-K^TK + C^TRC&=0.
\label{eq:met_smth_unc_ric2}
\end{align}
It will also include a solution to the differential equations:
\begin{equation}\label{eq:met_smth_unc_diff1}
\dot{\eta}(t) = -[A+B_1Q^{-1}B_1^TX]^T\eta(t)+C^TRy_0(t)+[K^TG+XB_2]u_0(t); \quad \eta(0) = X_0x_0
\end{equation}
for $t \in [0,\tau -q]$ and
\begin{equation}\label{eq:met_smth_unc_diff3}
-\dot{\xi}(t) = [A-B_1Q^{-1}B_1^TY]^T\xi(t)+C^TRy_0(t)-[YB_2-K^TG]u_0(t); \quad \xi(\tau) = 0
\end{equation}
for $t \in [\tau -q,\tau]$. Here, $y(t) = y_0(t)$ is a fixed measured output of the uncertain system (\ref{eq:met_smth_unc_4}), defined on the time interval $[0,\tau]$, and $u(t) = u_0(t)$ is a fixed measured input to the uncertain system defined on the same time interval.

Note that the form of (\ref{eq:met_smth_unc_diff3}) we are interested in is with respect to the running time variable $q$ that proceeds backward in time:
\begin{equation}\label{eq:met_smth_unc_diff2}
\dot{\xi}(q) = [A-B_1Q^{-1}B_1^TY]^T\xi(q)+C^TRy_0(q)-[YB_2-K^TG]u_0(q); \quad \xi(0) = 0
\end{equation}
for $q \in [0,\tau -t]$.

\begin{theorem} (See Theorem 5.1 in Ref. \cite{MSP}) 
Assume that (\ref{eq:met_smth_unc_ric1}) has a solution such that $X > 0$ and (\ref{eq:met_smth_unc_ric2}) has a solution such that $Y > 0$. Then, the set $X_{\tau -q}[x_0,u_0(\cdot)|_0^\tau ,y_0(\cdot)|_0^\tau ,d]$ of all possible states $x(\tau -q)$ at time $\tau -q$ for the uncertain system (\ref{eq:met_smth_unc_4}), where (\ref{eq:met_smth_unc_iqc}) is satisfied, is bounded and is given by:
\begin{equation}\label{eq:met_smth_unc_ellipse}
\begin{split}
X_{\tau -q}[x_0,u_0(\cdot)|_0^\tau ,y_0(\cdot)|_0^\tau ,d] &= \left\lbrace x_{\tau -q}: x_{\tau -q}^TXx_{\tau -q}-2x_{\tau -q}^T\eta(\tau -q) + h_{\tau -q}\right.\\
&\left.+ x_{\tau -q}^TYx_{\tau -q}-2x_{\tau -q}^T\xi(\tau -q) + s_{\tau -q} \leq d \right\rbrace
\end{split}
\end{equation}
where $\eta(t)$ and $\xi(t)$ are solutions to (\ref{eq:met_smth_unc_diff1}) and (\ref{eq:met_smth_unc_diff3}) and
\begin{equation}
\begin{split}
h_{\tau -q} &= x_0^TX_0x_0 + \int_0^{\tau -q}\left\{y_0(t)^TRy_0(t) - u_0(t)^TG^TGu_0(t)\right.\\
&\left.- \eta(t)^TB_1Q^{-1}B_1^T\eta(t)+2u_0(t)^TB_2\eta(t)\right\}dt,\\
s_{\tau -q} &= \int_{\tau -q}^\tau\left\{y_0(t)^TRy_0(t) - u_0(t)^TG^TGu_0(t)\right.\\
&\left.-\xi(t)^TB_1Q^{-1}B_1^T\xi(t)-2u_0(t)^TB_2\xi(t)\right\}dt.
\end{split}
\end{equation}
\end{theorem}

Clearly, the set of all possible states in (\ref{eq:met_smth_unc_ellipse}) is an ellipsoid, and the best estimate of the state is chosen as the centre of the ellipsoid.

\paragraph{Remark.} In the above, $\tilde{Q}$ and $\tilde{R}$ have been chosen to be arbitrarily close to $Q$ and $R$, respectively, which corresponds to small $\epsilon$. In the limit $\epsilon \to 0$, $d$ in (\ref{eq:met_smth_unc_12}), and therefore in (\ref{eq:met_smth_unc_iqc}), approaches infinity. However, in practice there will be a trade-off between how close $\tilde{Q}$ and $\tilde{R}$ are to $Q$ and $R$, respectively, and how large is $d$. The proposed theory is perfectly valid for finite values of $d$ in the IQC of (\ref{eq:met_smth_unc_iqc}), and it is never intended that the limit as $d$ approaches infinity should be considered.

Increasing the value of $d$ increases the diameter of the state estimation ellipse. However, the centre of the ellipse and hence the robust estimate is independent of $d$. Hence, in examples such as the one considered in this paper, in which we are only interested in the robust estimator (whose performance is validated via other means) and not the estimation ellipse, it would not be a problem if the value of $d$ chosen is very large (but finite).

\section{Standard Quantum Limit (SQL) for Resonant Noise}\label{sec:sql_resonant}
\begin{figure}[!t]
\centering
\includegraphics[width=0.9\textwidth]{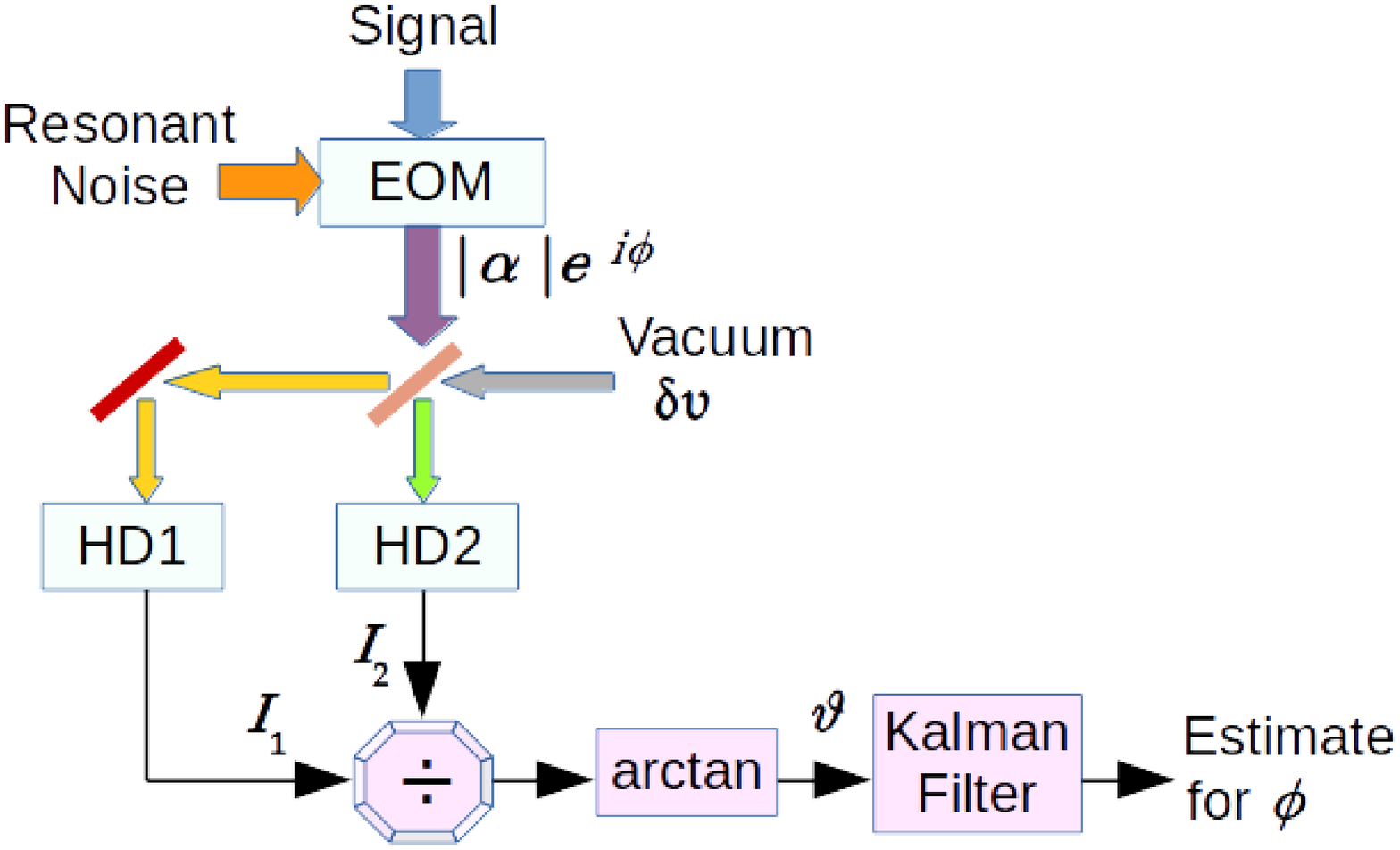}
\caption{Schematic diagram of optimal dual-homodyne phase estimation.}
\label{fig:dual_hd_sql}
\end{figure}

The standard quantum limit is set by the minimum error in phase estimation that can be obtained using a perfect heterodyne scheme with a coherent beam \cite{RPH2,RPH4}. We use the fact that the heterodyne scheme of measurement is, in principle, equivalent to, and incurs the same noise penalty as the \emph{dual-homodyne} scheme \cite{TW}, such as in the schematic depicted in Fig. \ref{fig:dual_hd_sql}. A coherent signal at the input is phase-modulated using an electro-optic modulator (EOM) that is driven by the resonant noise source. The modulated signal is then split using a $50-50$ beamsplitter into two arms each with a homodyne detector (HD1 and HD2, respectively, with the local oscillator phase of HD1 $\pi/2$ out of phase with that of HD2). The ratio of the output signals of the two arms goes to an \emph{arctan} block. The output of the arctan block is fed to an optimal Kalman filter, that yields the phase estimate with the minimum estimation error.

The output signals of the two arms are \cite{RPH2,RPH4}:
\begin{equation}
\begin{split}
I_1 &= \frac{1}{\sqrt{2}} \left( 2|\alpha| \sin\phi + \nu_1 + \nu_2 \right),\\ 
I_2 &= \frac{1}{\sqrt{2}} \left( 2|\alpha| \cos\phi + \nu_3 - \nu_4 \right),
\end{split}
\end{equation}
where $\nu_1$ and $\nu_3$ are measurement noises of the two homodyne detectors, respectively, and $\nu_2$ and $\nu_4$ are the noises arising from the vacuum entering the empty port of the input beamsplitter corresponding to the two arms, respectively. All these noises are assumed to be zero-mean white Gaussian noises.

The output of the arctan block is \cite{RPH2,RPH4}:
\begin{equation}
\vartheta = \arctan \left( \frac{2|\alpha|\sin\phi + \nu_1 + \nu_2}{2|\alpha|\cos\phi + \nu_3 - \nu_4} \right).
\end{equation}

Assuming the input noises are small, a Taylor series expansion up to first-order terms of the right-hand side yields \cite{RPH2,RPH4}:
\begin{equation}
\vartheta \approx \phi + \frac{1}{2|\alpha|}\nu_1 + \frac{1}{2|\alpha|}\nu_2.
\end{equation}

Expressing this equation in terms of $x$ in (\ref{eq:process_eqn}), we get the measurement model as \cite{RPH4}:
\begin{equation}\label{eq:dual_hd_meas}
\vartheta = Cx+D\nu,
\end{equation}
where $C = \left[\begin{array}{cc} 1 & 0 \end{array}\right]$, $D = \left[\begin{array}{cc} \frac{1}{2|\alpha|} & \frac{1}{2|\alpha|} \end{array}\right]$ and $\nu = \left[\begin{array}{c} \nu_1\\ \nu_2 \end{array}\right]$.

The error covariance matrix of the optimal steady-state Kalman filter for the process given by (\ref{eq:process_eqn}) and the measurement given by (\ref{eq:dual_hd_meas}) may be obtained by solving an algebraic Riccati equation of the form (\ref{eq:met_kalman_alg_riccati}) for $P$. The error covariance of interest (i.e. that in estimating $\phi$) is then $\sigma^2 = P(1,1)$.

\section*{References}
\bibliography{bibliography.bib}

\providecommand{\newblock}{}
\begin{thebibliography}{10}
\expandafter\ifx\csname url\endcsname\relax
  \def\url#1{{\tt #1}}\fi
\expandafter\ifx\csname urlprefix\endcsname\relax\def\urlprefix{URL }\fi
\providecommand{\eprint}[2][]{\url{#2}}

\bibitem{WM}
Wiseman H~M and Milburn G~J 2010 {\em Quantum Measurement and Control\/}
  (Cambridge University Press)

\bibitem{HWA}
Hofheinz M, Wang H, Ansmann M, Bialczak R~C, Lucero E, Neeley M, O'Connell A~D,
  Sank D, Wenner J, Martinis J~M and Cleland A~N 2009 {\em Nature (London)\/}
  {\bf 459} 546--549

\bibitem{SPK}
Slavik R, Parmigiani F, Kakande J, Lundstrom C, Sjodin M, Andrekson P~A,
  Weerasuriya R, Sygletos S, Ellis A~D, Gruner-Nielsen L, Jakobsen D, Herstrom
  S, Phelan R, O'Gorman J, Bogris A, Syvridis D, Dasgupta S, Petropoulos P and
  Richardson D~J 2010 {\em Nature Photonics\/} {\bf 4} 690--695

\bibitem{CHD}
Chen J, Habif J~L, Dutton Z, Lazarus R and Guha S 2012 {\em Nature Photonics\/}
  {\bf 6} 374

\bibitem{IWY}
Inoue K, Waks E and Yamamoto Y 2002 {\em Physical Review Letters\/} {\bf 89}
  037902

\bibitem{GLM1}
Giovannetti V, Lloyd S and Maccone L 2011 {\em Nature Photonics\/} {\bf 5} 222

\bibitem{GMM}
Goda K, Miyakawa O, Mikhailov E~E, Saraf S, Adhikari R, McKenzie K, Ward R,
  Vass S, Weinstein A~J and Mavalvala N 2008 {\em Nature Physics\/} {\bf 4}
  472--476

\bibitem{TW}
Wheatley T~A, Berry D~W, Yonezawa H, Nakane D, Arao H, Pope D~T, Ralph T~C,
  Wiseman H~M, Furusawa A and Huntington E~H 2010 {\em Physical Review
  Letters\/} {\bf 104} 093601

\bibitem{YNW}
Yonezawa H, Nakane D, Wheatley T~A, Iwasawa K, Takeda S, Arao H, Ohki K,
  Tsumura K, Berry D~W, Ralph T~C, Wiseman H~M, Huntington E~H and Furusawa A
  2012 {\em Science\/} {\bf 337} 1514

\bibitem{LXP}
Lewis F~L, Xie L and Popa D 2008 {\em Optimal and Robust Estimation - With an
  Introduction to Stochastic Control Theory\/} 2nd ed (CRC Press, Taylor \&
  Francis Group)

\bibitem{ZDG}
Zhou K, Doyle J~C and Glover K 1996 {\em Robust and Optimal Control\/}
  (Prentice-Hall)

\bibitem{JS}
Stockton J~K, Geremia J~M, Doherty A~C and Mabuchi H 2004 {\em Physical Review
  A\/} {\bf 69} 032109

\bibitem{NY}
Yamamoto N 2006 {\em Physical Review A\/} {\bf 74} 03217

\bibitem{JNP}
James M~R, Nurdin H~I and Petersen I~R 2008 {\em IEEE Trans. on Automatic
  Control\/} {\bf 53} 1787

\bibitem{HMW}
Wiseman H~M 1995 {\em Physical Review Letters\/} {\bf 75} 4587--4590

\bibitem{WK1}
Wiseman H~M and Killip R~B 1997 {\em Physical Review A\/} {\bf 56} 944--957

\bibitem{WK2}
Wiseman H~M and Killip R~B 1998 {\em Physical Review A\/} {\bf 57} 2169--2185

\bibitem{PWL}
Pope D~T, Wiseman H~M and Langford N~K 2004 {\em Physical Review A\/} {\bf 70}
  043812

\bibitem{MA}
Armen M~A, Au J~K, Stockton J~K, Doherty A~C and Mabuchi H 2002 {\em Physical
  Review Letters\/} {\bf 89} 133602

\bibitem{BW1}
Berry D~W and Wiseman H~M 2000 {\em Physical Review A\/} {\bf 63} 013813

\bibitem{BW2}
Berry D~W and Wiseman H~M 2002 {\em Physical Review A\/} {\bf 65} 043803

\bibitem{TSL}
Tsang M, Shapiro J~H and Lloyd S 2009 {\em Physical Review A\/} {\bf 79} 053843

\bibitem{MT1}
Tsang M 2009 {\em Physical Review Letters\/} {\bf 102} 250403

\bibitem{LK}
Ljung L and Kailath T 1976 {\em Automatica\/} {\bf 12} 147--157

\bibitem{JSM}
Meditch J~S 1973 {\em Automatica\/} {\bf 9} 151--162

\bibitem{FP}
Fraser D~C and Potter J~E 1969 {\em IEEE Transactions on Automatic Control\/}
  {\bf 14} 387--390

\bibitem{WWS}
{Wall Jr} J~E, Willsky A~S and {Sandell Jr} N~R 1981 {\em Stochastics\/} {\bf
  5} 1--41

\bibitem{DQM}
Mayne D~Q 1966 {\em Automatica\/} {\bf 4} 73--92

\bibitem{DCF}
Fraser D~C 1967 {\em A New Technique for the Optimal Smoothing of Data\/}
  Sc.{D}. dissertation Massachusetts Institute of Technology, Cambridge, MA

\bibitem{RKM}
Mehra R~K 1967 {\em Studies in Smoothing and in Conjugate Gradient Methods
  Applied to Optimal Control Problems\/} Ph.{D}. dissertation Harvard
  University, Cambridge, MA

\bibitem{MSP}
Moheimani S~O~R, Savkin A~V and Petersen I~R 1998 {\em IEEE Trans. on Circuits
  and Systems I - Fundamental Theory and Appl.\/} {\bf 45} 446

\bibitem{RPH2}
Roy S, Petersen I~R and Huntington E~H 2013 Adaptive continuous homodyne phase
  estimation using robust fixed-interval smoothing {\em Proceedings of the
  American Control Conference\/} pp 3129--3134

\bibitem{RPH3}
Roy S, Petersen I~R and Huntington E~H 2013 Robust phase estimation of squeezed
  state {\em Proceedings of the Conference on Lasers and Electro-Optics\/} p
  JTh2A.88

\bibitem{RPH5}
Roy S, Rehman O, Petersen I~R and Huntington E~H 2014 Robust smoothing for
  estimating optical phase varying as a continuous resonant process {\em
  Proceedings of the European Control Conference\/} pp 896--901

\bibitem{RPH1}
Roy S, Petersen I~R and Huntington E~H 2012 Robust filtering for adaptive
  homodyne estimation of continuously varying optical phase {\em Proceedings of
  the Australian Control Conference\/} pp 454--458

\bibitem{RPH4}
Roy S, Petersen I~R and Huntington E~H 2013 Robust estimation of optical phase
  varying as a continuous resonant process {\em Proceedings of the
  Multiconference on Systems and Control\/} pp 551--555

\bibitem{RGB}
Brown R~G and Hwang P~Y~C 1997 {\em Introduction to Random Signals and Applied
  Kalman Filtering\/} (John Wiley \& Sons) pp 289--293,322--325 3rd ed

\bibitem{KI}
Iwasawa K, Makino K, Yonezawa H, Tsang M, Davidovic A, Huntington E and
  Furusawa A 2013 {\em Physical Review Letters\/} {\bf 111} 163602

\end{thebibliography}

\end{document}